\documentclass[aps,preprint,epsf]{revtex4}
\usepackage{amsmath}
\usepackage{amsfonts}
\usepackage{graphicx,graphics}

\input{tcilatex}

\begin{document}

\title{Functional renormalization-group approaches, one-particle
(ir)reducible with respect to local Green functions, using the dynamical
mean-field theory as a starting point}
\author{A. A. Katanin}

\begin{abstract}
We consider formulations of the functional renormaliztion-group flow for
correlated electronic systems, having the dynamical mean-field theory as a
starting point. We classify the corresponding renormalization-group schemes
into those neglecting the one-particle irreducible (with respect to the
local Green functions) six-point vertices and neglecting one-particle
reducible six-point vertices. The former class is represented by the
recently introduced DMF$^{2}$RG approach [Phys. Rev. Lett. \textbf{112},
196402 (2014)], but also by the scale-dependent generalization of the
one-particle irreducible (with respect to local Green functions, 1PI-LGF)
representation of the generating functional [Phys. Rev. B \textbf{88},
115112 (2013)]. The second class is represented by the fRG flow within the
dual fermion (DF) approach [Phys. Rev. B \textbf{77}, 033101 (2008); ArXiv
1411.1342]. We compare formulations of fRG approach in each of these cases
and suggest their further application to study 2D systems within the Hubbard
model.
\end{abstract}

\maketitle

\address{Institute of Metal Physics, 620990, Ekaterinburg, Russia\\
Ural Federal University, 620002 Ekaterinburg, Russia}

%\address{$^a$ Max-Planck-Institut f\"ur Festk\"orperforschung, D-70569,\\
%Stuttgart\\
%$^b$ Institute of Metal Physics, 620219 Ekaterinburg, Russia}

\section{Introduction}

Strongly correlated electron systems demonstrate a variety of interesting
phenomena, such as magnetism, (unconventional) superconductivity,
\textquotedblleft colossal\textquotedblright\ magnetoresistance, and quantum
critical behavior. The dynamical mean-field theory (DMFT)\cite{DMFT,DMFT2},
which becomes exact in the limit of high spatial dimensions ($d\rightarrow
\infty $),\ allowed to achieve a substantial progress in describing strong
electronic correlations. In particular it allowed to describe accurately the
Mott-Hubbard metal-insulator transition\cite{MH} due to account of an
important local part of electronic correlations.

In real physical systems, which are one-, two-, or three-dimensional, the
nonlocal correlations, which are neglected in DMFT, are, however, important.
Cluster extensions of DMFT \cite%
{clusterDMFT,clusterDMFT1,clusterDMFT2,clusterDMFT3,clusterDMFT4} can treat
only short-range correlations due to numerical limitations of the cluster
size\cite{clusterDMFTFlex}. In spite of this, the diagrammatic extensions of
the dynamical mean-field theory were developed. These are the dynamical
vertex approximation (D$\Gamma $A) \cite%
{DGA1a,DGA1b,DGA1c,DGA1d,DGA2,Kusunose}, the dual fermion (DF) approach \cite%
{DF1,DF2,DF3,DF4}, and the one-particle irreducible with respect to the
local Green functions (1PI-LGF) approach \cite{Our1PI}. The former
approximation starts from the local two-particle irreducible vertices and
sums ladder or parquet diagrams for the vertex, considering the effect of
the non-locality of the Green functions. The DF approach on the other hand
splits the degrees of freedom into the local ones, treated within DMFT, and
the non-local (dual) degrees, considered perturbatively, with a possibility
of summation of infinite series of diagrams for dual fermions \cite%
{DF4,Jarrell}. The 1PI version of the dual fermion approach (the 1PI-LGF
approach) performs the same splitting of the local and non-local degrees of
freedom for 1PI (Legendre-transformed) generating functionals. This approach
accounts therefore for the effect of one-particle reducible six-point and
higher-order reducible vertices, which was argued to be of possible
importance in Ref. \cite{MySixpt}.

The abovediscussed approaches typically treat non-local fluctuations within
the ladder approximation. More powerful method - parquet approach can bring
substantial improvement over the ladder approximation\cite{Jarrell,Toschi1},
but often not accessible numerically for correlated electronic systems. At
the same time, recently developed functional renormalization group (fRG)
approaches \cite{Salmhofer,Zanchi,Metzner,Honerkamp,Katanin,fRGReview} allow
for lower computational cost perform approximate summation of the parquet
set of the diagrams, provided that the six-point (i.e. three-particle)
interaction vertices remain sufficiently small during the flow. In
particular, for the standard fRG, applied to the Hubbard model, the initial
particle irreducible six-point vertices are zero, which advances using the
one-particle irreducible approach, in which one can expect that the
corresponding six-point vertex remain small during the flow. Using the
dynamical mean-field theory as an initial theory for the flow account
exactly for the local subset of diagrams, but yields (in principle) non-zero
vertices up to infinite order, so that the formulation and justification of
the fRG approach requires more effort. In general, in this case, one has a
choice between neglecting the six-point one-particle \textit{irreducibe }or%
\textit{\ reducible }vertices, which depends on considering model. In
particular, in the half-filled spinless Falikov-Kimaball model the
one-particle reducible six-point local vertex vanishes in the
infinite-dimensional limit \cite{FK}, while for the Hubbard model, at least
in the weak-to-intermediate coupling limit, neglecting six-point
one-particle irreducible local vertices seems more preferable.

In the present paper we concentrate on the renormalization-group approaches,
which use dynamical mean-field theory as a starting point, and neglect
either one-particle irreducible or one-particle reducible (with respect to
the local Green function) three-particle vertices (see also Ref. \cite{OurRG}%
). Recently, an approach of the former type, considering the functional
renormalization-group flow from infinite to finite dimensions (the DMF$^{2}$%
RG approach) was introduced\cite{ifflow}. This flow starts from
infinite-dimensional model, which is solved by DMFT, and considers the flow
to finite number of dimensions, e.g. in the approximation of neglecting the
local six-point vertices. Because of the using 1PI approach, the latter
approximation implies neglect of the six-point vertices, which are
one-particle \textit{irreducible} with respect to the local Green functions.
To have more general view on the possible variety of different
renormalization-group approaches, starting from the dynamical mean-field
theory, it is informative to formulate the fRG approach for the other two
mentioned schemes, i.e. 1PI-LGF and DF theories, and compare them to the DMF$%
^{2}$RG approach. This study is performed in the present paper.

\section{The model and dynamical mean-field theory}

We consider general one-band model of fermions, interacting via local
interaction $H_{\mathrm{int}}[\widehat{c}_{i\sigma },\widehat{c}_{i\sigma
}^{+}]$ 
\begin{equation}
H=\sum\limits_{\mathbf{k},\sigma }\varepsilon _{\mathbf{k},\sigma }\widehat{c%
}_{\mathbf{k},\sigma }^{+}\widehat{c}_{\mathbf{k},\sigma
}+\sum\limits_{i,\sigma }H_{\mathrm{int}}[\widehat{c}_{i\sigma },\widehat{c}%
_{i\sigma }^{+}],  \label{H}
\end{equation}%
where $\widehat{c}_{i\sigma },\widehat{c}_{i\sigma }^{+}$ are the fermionic
operators, and $\widehat{c}_{\mathbf{k},\sigma },\widehat{c}_{\mathbf{k}%
,\sigma }^{+}$ are the corresponding Fourier transformed objects, $\sigma
=\uparrow ,\downarrow $ corresponds to a spin index. The model is
characterized by the generating functional%
\begin{eqnarray}
Z[\eta ,\eta ^{+}] &=&\int d[c,c^{+}]\exp \left\{ -\mathcal{S}[c,c^{+}]+\eta
^{+}c+c^{+}\eta \right\}  \label{gen} \\
\mathcal{S}[c,c^{+}] &=&\int d\tau \left\{ \sum\limits_{i,\sigma }c_{i\sigma
}^{\dagger }(\tau )\frac{\partial }{\partial \tau }c_{i\sigma }(\tau
)+H[c,c^{+}]\right\}
\end{eqnarray}%
where $c_{i\sigma },c_{i\sigma }^{+},\eta _{i\sigma },\eta _{i\sigma }^{+}$
are the Grassman fields, the fields $\eta _{i\sigma },\eta _{i\sigma }^{+}$
correspond to source terms, $\tau \in \lbrack 0,\beta =1/T]$ is the
imaginary time. The dynamical mean-field theory \cite{DMFT,DMFT2} for the
model (\ref{H}) can be introduced via effecive interaction 
\begin{eqnarray}
\mathcal{V}_{\mathrm{DMFT}}[\eta ,\eta ^{+}] &=&-\ln \int d[c,c^{+}]\exp
\left\{ -\sum\limits_{i,\sigma }\int d\tau H_{\mathrm{int}}[c_{i\sigma
},c_{i\sigma }^{+}]\right.  \notag \\
&&\left. +\sum\limits_{k,\sigma }\zeta ^{-1}(i\nu _{n})\left( c_{k,\sigma
}^{+}+\eta _{k,\sigma }^{+})(c_{k,\sigma }+\eta _{k,\sigma }\right) \right\}
\label{VDMFT}
\end{eqnarray}%
where the "Weiss field" function $\zeta (\tau )$ and its Fourier transform $%
\zeta (i\nu _{n})$ has to be determined self-consistently from the condition%
\begin{equation}
G_{\mathrm{loc}}(i\nu _{n})\equiv \frac{1}{\zeta ^{-1}(i\nu _{n})-\Sigma _{%
\mathrm{loc}}(i\nu _{n})}=\sum\limits_{\mathbf{k}}\mathcal{G}(\mathbf{k}%
,i\nu _{n}),  \label{sc}
\end{equation}%
where%
\begin{equation}
\mathcal{G}(\mathbf{k},i\nu _{n})\equiv \mathcal{G}_{k}=\left[
G_{0k}^{-1}-\Sigma _{\mathrm{loc}}(i\nu _{n})\right] ^{-1},  \label{Glattice}
\end{equation}%
$G_{0k}^{-1}=i\nu _{n}-\varepsilon _{\mathbf{k}}\,$is the lattice
noninteracting Green function (we use the $4$-vector notation $k=(\mathbf{k}%
,i\nu _{n})$) and $\Sigma _{\mathrm{loc}}(i\nu _{n})$ is the self-energy of
the impurity problem (\ref{VDMFT}), which is in practice obtained within one
of the impurity solvers: exact diagonalization, quantum Monte-Carlo, etc.
These solvers provide information not only on the electronic self-energy,
but also the corresponding vertex functions\cite{DGA1a,Toschi}. This is
reflected in the following expansion of the effective interaction: 
\begin{equation}
\mathcal{V}_{\mathrm{DMFT}}[\eta ,\eta ^{+}]=\widehat{\mathcal{V}}_{\mathrm{%
DMFT}}[\widehat{\eta }_{k\sigma },\widehat{\eta }_{k\sigma
}^{+}]+\sum\limits_{k,\sigma }\eta _{k,\sigma }^{+}\frac{\Sigma _{\mathrm{loc%
}}(i\nu _{n})}{1-\zeta (i\nu _{n})\Sigma _{\mathrm{loc}}(i\nu _{n})}\eta
_{k,\sigma },
\end{equation}%
where $\widehat{\eta }_{k,\sigma }=\eta _{k\sigma }/(1-\Sigma _{\mathrm{loc}%
}(i\nu _{n})G_{0k})$. The functional $\widehat{\mathcal{V}}_{\mathrm{DMFT}%
}[\eta ,\eta ^{+}]$ generates connected vertices (which are in general
one-particle reducible), amputated by the local Green function $G_{\mathrm{%
loc}}(i\nu _{n})$, such that its expansion in fields reads 
\begin{eqnarray}
\widehat{\mathcal{V}}_{\mathrm{DMFT}}[\eta ^{+},\eta ] &=&\frac{1}{2}\Gamma
_{\mathrm{loc}}\circ (\eta _{k_{1},\sigma }^{+}\eta _{k_{3},\sigma })(\eta
_{k_{2},\sigma ^{\prime }}^{+}\eta _{k_{4},\sigma ^{\prime }})
\label{VDMFT1} \\
&&+\frac{1}{6}\Gamma _{\mathrm{loc}}^{(6)}\circ (\eta _{k_{1\sigma
}}^{+}\eta _{k_{2\sigma }})(\eta _{k_{3}\sigma ^{\prime }}^{+}\eta
_{k_{4}\sigma ^{\prime }})(\eta _{k_{5}\sigma ^{\prime \prime }}^{+}\eta
_{k_{6}\sigma ^{\prime \prime }})+...  \notag
\end{eqnarray}%
where $\Gamma _{\mathrm{loc}}$ and $\Gamma _{\mathrm{loc}}^{(6)}$ are the
connected $4$- and $6$-point vertices, amputated with the local Green
functions $G_{\mathrm{loc}},$ e.g. 
\begin{eqnarray}
\widetilde{\Gamma }_{\mathrm{loc}}^{\sigma \sigma ^{\prime }}(i\nu
_{1}..i\nu _{3}) &=&(1+\delta _{\sigma \sigma ^{\prime
}})^{-1}\prod\nolimits_{i=1}^{4}G_{\mathrm{loc}}^{-1}(i\nu _{i})
\label{Gamma_loc} \\
&&\times \left[ G_{\mathrm{loc,}\sigma \sigma ^{\prime }}^{(4)}(i\nu
_{1}..i\nu _{3})-G_{\mathrm{loc}}(i\nu _{1})G_{\mathrm{loc}}(i\nu
_{2})(\delta _{\nu _{1}\nu _{3}}-\delta _{\sigma \sigma ^{\prime }}\delta
_{\nu _{2}\nu _{3}})\right] ,  \notag
\end{eqnarray}%
and $\circ $ stands for summation over momenta- frequency- and spin indices
fulfilling the conservation laws, $G_{\mathrm{loc}}^{(4)}$ is the
two-particle local Green function, which can be obtained via the solution of
the impurity problem. For the four-point vertex $\Gamma _{\mathrm{loc}}$ the
requirement of connectivity and amputation with the full local Green
functions implies one-particle irreducibility. However, the higher-order
vertices, e.g. $\Gamma _{\mathrm{loc}}^{(6)}$ remain one-particle reducible
with respect to the local Green functions. To obtain the one-particle
irreducible vertices, Legendre transformation of Eq. (\ref{VDMFT}) has to be
performed.

\section{The one-particle irreducible approaches with respect to local Green
functions}

\subsection{The flow from infinite to finite dimension within the DMF$^{2}$%
RG approach}

Recently, in Ref. \cite{ifflow} the flow from infinite to finite number of
dimensions was introduced. This flow considers evolution of generating
functional with the action 
\begin{equation}
\mathcal{S}_{\Lambda }=\sum_{k\sigma }c_{k\sigma }^{+}G_{0k,\Lambda
}^{-1}c_{k\sigma }+\sum\limits_{i,\sigma }\int d\tau H_{\mathrm{int}%
}[c_{i\sigma },c_{i\sigma }^{+}]  \label{eq:S_L}
\end{equation}%
with the cutoff dependence of the bare Green function%
\begin{equation}
G_{0k,\Lambda }=1/[f(k,\Lambda )G_{0,k}^{-1}+(1-f(k,\Lambda ))\zeta
^{-1}(i\nu _{n})]  \label{G_L_ifflow0}
\end{equation}%
with some function $f(k,\Lambda ),$ such that $f(k,1)=0$ and $f(k,0)=1;$
specific choices of this function are discussed in Sect. IV ($f(k,\Lambda
)=1-\Lambda $ was used in Ref. \cite{ifflow}). The 1PI approach, applied to
the model (\ref{eq:S_L}) yields equations 
\begin{eqnarray}
\frac{d\Sigma _{\Lambda }}{d\Lambda } &=&V_{\Lambda }\circ S_{\Lambda },
\label{Sflow_eq} \\
\frac{dV_{\Lambda }}{d\Lambda } &=&V_{\Lambda }\circ (G_{\Lambda }\circ
S_{\Lambda }+S_{\Lambda }\circ G_{\Lambda })\circ V_{\Lambda },
\label{ifflow_eq}
\end{eqnarray}%
where 
\begin{eqnarray}
G_{k,\Lambda } &=&G_{0k,\Lambda }/[1-\Sigma (k,\Lambda )G_{0k,\Lambda }],
\label{G_iffl1} \\
S_{k,\Lambda } &=&\left. \frac{dG_{k,\Lambda }}{d\Lambda }\right\vert
_{\Sigma =\mathrm{const}}=-(\mathcal{G}_{k}^{-1}-G_{\mathrm{loc}}^{-1})\frac{%
\partial f}{\partial \Lambda }G_{k,\Lambda }^{2}.  \label{SkL_iffl}
\end{eqnarray}%
This approach uses initial one-particle irreducible vertices and self-energy
as an initial condition for the flow: $V_{\Lambda =1}=\Gamma _{\mathrm{loc}%
}^{\uparrow \downarrow },$ $\Sigma _{\Lambda =1}=\Sigma _{\mathrm{loc}}$,
which are in practice obtained from the solution of the AIM. At the same
time, equations (\ref{Sflow_eq}) and (\ref{ifflow_eq}) neglect local 1PI
six-point vertex at the initial stage of the flow.

\subsection{General formulation of the 1PI-LGF approach}

Another way of the treatment of non-local correlations, based on the
dynamical mean-field theory as a starting point, is splitting of the local
and non-local correlations in the generating functional for lattice theory.
For one-particle irreducible version, this was done within the 1PI-LGF
approach, considered in Ref. \cite{Our1PI}. This approach represents the
partition function as a functional of the local Green function $G_{\mathrm{%
loc}}$ and the corresponding non-local part $\widetilde{\mathcal{G}}_{k}=%
\mathcal{G}_{k}-G_{\mathrm{loc}}$. Contrary to the dual fermion approach,
considered in the next Section, this representation contains two fermionic
fields, one of which describes propagation of non-local degrees of freedom
(similarly to the DF approach), while the other one provides one-particle
irreducibility of the resulting functional.

To formulate the renormalization-group treatment within this approach, we
generalize trivially the representation for the partition function, obtained
in Ref. \cite{Our1PI}, to introduce $\Lambda $-dependence of the lattice
Green function $\mathcal{G}_{k}$ by the replacement $\mathcal{G}%
_{k}\rightarrow \mathcal{G}_{k,\Lambda }$ where $\mathcal{G}_{k,\Lambda }$
is defined by 
\begin{equation}
\mathcal{G}_{k,\Lambda }=1/[f(k,\Lambda )\mathcal{G}_{k}^{-1}+(1-f(k,\Lambda
))G_{\mathrm{loc}}^{-1}(i\nu _{n})],  \label{G_L_ifflow}
\end{equation}%
which is similar to the Eq. (\ref{G_L_ifflow0}). The other choice, which we
consider below is combining the two Green functions (and not their inverse)
in a sum, 
\begin{equation}
\mathcal{G}_{k,\Lambda }=f(k,\Lambda )\mathcal{G}_{k}+[1-f(k,\Lambda )]G_{%
\mathrm{loc}}(i\nu _{n})  \label{G_L_sum}
\end{equation}%
such that 
\begin{equation}
\widetilde{\mathcal{G}}_{k}\rightarrow \widetilde{\mathcal{G}}_{k,\Lambda }:=%
\mathcal{G}_{k,\Lambda }-G_{\mathrm{loc}}(i\nu _{n})=f(k,\Lambda )[\mathcal{G%
}_{k}-G_{\mathrm{loc}}(i\nu _{n})].  \label{G_tilde}
\end{equation}%
The resulting $\Lambda $-dependent partition function in both cases reads%
\cite{Our1PI}:%
\begin{eqnarray}
Z_{\Lambda }[\eta ^{+},\eta ] &=&\int D[\phi ^{+},\phi ]D[\psi ^{+},\psi
]\;\exp \left\{ \sum_{k,\sigma }\eta _{k\sigma }^{+}\left( \psi _{k\sigma
}+\phi _{k\sigma }\right) +\left( \psi _{k\sigma }^{+}+\phi _{k\sigma
}^{+}\right) \eta _{k\sigma }\right.  \notag \\
\hspace{0.7cm} &+&\left. \frac{1}{\beta }\sum_{k,\sigma }\mathcal{G}%
_{k,\Lambda }^{-1}\left( \phi _{k\sigma }^{+}\phi _{k\sigma }+\psi _{k\sigma
}^{+}\phi _{k\sigma }+\phi _{k\sigma }^{+}\psi _{k\sigma }\right) +\left( 
\mathcal{G}_{k,\Lambda }^{-1}-G_{\mathrm{loc},\nu }^{-1}\right) \psi
_{k\sigma }^{+}\psi _{k\sigma }\right.  \notag \\
\hspace{1.9cm} &-&\frac{1}{\beta ^{3}}\sum_{kk^{\prime }q}\sum_{\sigma
\sigma ^{\prime }}\widetilde{\Gamma }_{\mathrm{loc},\sigma \sigma ^{\prime
}}^{\nu \nu ^{\prime }\omega }\left[ \left( \psi _{k\sigma }^{+}\phi
_{k+q,\sigma }\right) \left( \phi _{k^{\prime }+q,\sigma ^{\prime }}^{+}\phi
_{k^{\prime }\sigma ^{\prime }}\right) \right.  \notag \\
&+&\left. \left( \phi _{k\sigma }^{+}\phi _{k+q,\sigma }\right) \left( \phi
_{k^{\prime }+q,\sigma ^{\prime }}^{+}\psi _{k^{\prime }\sigma ^{\prime
}}\right) +\frac{1}{2}\left( \phi _{k\sigma }^{+}\phi _{k+q,\sigma }\right)
\left( \phi _{k^{\prime }+q,\sigma ^{\prime }}^{+}\phi _{k^{\prime }\sigma
^{\prime }}\right) \bigr]\right\} J[\phi ^{+},\phi ],  \label{Zf}
\end{eqnarray}%
where $J[\phi ^{+},\phi ]$ is the Jacobian, defined in terms of local
degrees of freedom in Ref. \cite{Our1PI}; $\widetilde{\Gamma }_{\mathrm{loc}%
,\sigma \sigma ^{\prime }}^{\nu \nu ^{\prime }\omega }=\left( 1-\delta
_{\sigma \sigma ^{\prime }}/2\right) \Gamma _{\mathrm{loc},\sigma \sigma
^{\prime }}^{\nu \nu ^{\prime }\omega }$. Eq. (\ref{Zf}) contains
integration over two fermionic fields $\phi $ and $\psi ,$ the latter
appears after fermionic Hubbard-Stratanovich transformation of the Legandre
transformation of the action and provides one-particle irreducibility of the
resulting approach with respect to the local Green functions. The
diagrammatic meaning of Eq. (\ref{Zf}), as well as the summation of the
ladder diagrams for the vertex and their effect on the self-energy was
discussed in details in Ref. \cite{Our1PI}; here we consider
renormalization-group approach to this representation.

The bare propagator of the representation (\ref{Zf}), which includes fully
the effect of the local self-energy, can be conveniently written in the
spinor representation\cite{Our1PI}, 
\begin{equation}
\Phi _{k\sigma }=\left( 
\begin{tabular}{l}
$\phi _{k\sigma }$ \\ 
$\psi _{k\sigma }$%
\end{tabular}%
\ \right) ,
\end{equation}%
and it reads 
\begin{equation}
\mathbf{G}_{k,\Lambda }=-\frac{1}{\beta }\langle \langle \Phi _{k}|\Phi
_{k}^{+}\rangle \rangle _{0}=\left( 
\begin{array}{cc}
\mathcal{G}_{k,\Lambda }^{-1} & \mathcal{G}_{k,\Lambda }^{-1} \\ 
\mathcal{G}_{k,\Lambda }^{-1} & \mathcal{G}_{k,\Lambda }^{-1}-G_{\mathrm{loc}%
,\nu }^{-1}%
\end{array}%
\right) ^{-1}=\left( 
\begin{array}{cc}
\widetilde{\mathcal{G}}_{k,\Lambda } & G_{\mathrm{loc},\nu } \\ 
G_{\mathrm{loc},\nu } & -G_{\mathrm{loc},\nu }%
\end{array}%
\right) .  \label{G_nonint}
\end{equation}%
The corresponding equations for the vertex $\mathbb{V}_{\Lambda }^{\alpha
\beta \gamma \delta }(k_{1},k_{2};k_{3},k_{4})$ ($k_{1},k_{2}$ and $%
k_{3},k_{4}$ are the momenta- and frequencies of the incoming and outgoing
electrons, $k_{i}=(\mathbf{k}_{i},i\nu _{n}^{(i)}),$ $\alpha ,\beta ,\gamma
,\delta =1,2$ correspond to $\phi $ and $\psi $ fields, respectively) and
the non-local part of the self-energy $\widetilde{\Sigma }_{\Lambda
}^{\alpha \beta }(\mathbf{k},i\nu _{n})$ read 
\begin{eqnarray}
\frac{d\widetilde{\Sigma }_{\Lambda }}{d\Lambda } &=&\mathbb{V}_{\Lambda
}\circ \mathbb{S}_{\Lambda };  \label{OneLoopA} \\
\frac{d\mathbb{V}_{\Lambda }}{d\Lambda } &=&\mathbb{V}_{\Lambda }\circ (%
\mathbb{G}_{\Lambda }\circ \mathbb{S}_{\Lambda }+\mathbb{S}_{\Lambda }\circ 
\mathbb{G}_{\Lambda })\circ \mathbb{V}_{\Lambda },  \label{OneLoop}
\end{eqnarray}%
where 
\begin{equation}
\mathbb{G}_{k,\Lambda }=[\mathbf{G}_{k,\Lambda }^{-1}-\widetilde{\mathbf{%
\Sigma }}_{k,\Lambda }]^{-1}.
\end{equation}%
For the choice of the propagators (\ref{G_L_ifflow}) we obtain 
\begin{equation}
\mathbb{S}_{k,\Lambda }=-(\mathcal{G}_{k}^{-1}-G_{\mathrm{loc}}^{-1})\frac{%
\partial f(k,\Lambda )}{\partial \Lambda }[\mathbf{G}_{k,\Lambda }^{-1}-%
\widetilde{\mathbf{\Sigma }}_{k,\Lambda }]^{-1}\left( 
\begin{array}{cc}
1 & 1 \\ 
1 & 1%
\end{array}%
\right) [\mathbf{G}_{k,\Lambda }^{-1}-\widetilde{\mathbf{\Sigma }}%
_{k,\Lambda }]^{-1},  \label{S_1PI_Metz}
\end{equation}%
while for the propagator (\ref{G_L_sum}) we find 
\begin{equation}
\mathbb{S}_{k,\Lambda }=-(\mathcal{G}_{k}-G_{\mathrm{loc}})\mathcal{G}%
_{k,\Lambda }^{-2}\frac{\partial f(k,\Lambda )}{\partial \Lambda }[\mathbf{G}%
_{k,\Lambda }^{-1}-\widetilde{\mathbf{\Sigma }}_{k,\Lambda }]^{-1}\left( 
\begin{array}{cc}
1 & 1 \\ 
1 & 1%
\end{array}%
\right) [\mathbf{G}_{k,\Lambda }^{-1}-\widetilde{\mathbf{\Sigma }}%
_{k,\Lambda }]^{-1}.  \label{S_1PI_Sum}
\end{equation}%
Note that the non-local (physical) Green function can be directly obtained
from Eqs. (\ref{OneLoop}) by summing all the components of the matrix Green
function $(\mathbf{G}_{k,\Lambda }^{-1}-\widetilde{\mathbf{\Sigma }}%
_{k,\Lambda })^{-1}$, the corresponding `physical' self-energy is then
extracted in the standard way from the physical Green function and the
flowing bare Green function $(\mathcal{G}_{k,\Lambda }^{-1}+\Sigma _{\mathrm{%
loc}})^{-1}$.

For $\Lambda \geq \Lambda _{0}$ ($\Lambda _{0}$ is the upper scale of the
problem) we have $\widetilde{G}_{\Lambda }=0$, so that only $\mathbf{G}^{12},%
\mathbf{G}^{21},$ and $\mathbf{G}^{22}$ elements of the Green function are
nonzero, which corresponds to a purely local theory. It can be shown that
the contribution of these Green functions are exactly compensated by the
"counterterms" which arise from the Jacobian of the transformation. The
initial conditions for the vertex and the self-energy are 
\begin{eqnarray}
\mathbb{V}_{\Lambda }^{1111}(k_{1},k_{2};k_{3},k_{4}) &=&\mathbb{V}_{\Lambda
}^{1211}(k_{1},k_{2};k_{3},k_{4})=\mathbb{V}_{\Lambda
}^{2111}(k_{1},k_{2};k_{3},k_{4})  \notag \\
&=&\mathbb{V}_{\Lambda }^{1121}(k_{1},k_{2};k_{3},k_{4})=\mathbb{V}_{\Lambda
}^{1112}(k_{1},k_{2};k_{3},k_{4})  \label{V_struct} \\
&=&\Gamma _{\mathrm{loc}}^{\uparrow \downarrow },\   \notag \\
\widetilde{\mathbf{\Sigma }}^{\alpha \beta }(k) &=&0.
\end{eqnarray}

\subsection{Comparison to the DMF$^{2}$RG approach}

To compare the fRG flow within 1PI-LGF and DMF$^{2}$RG approaches, we
consider the choice of the propagators (\ref{G_L_ifflow}) and (\ref%
{S_1PI_Metz}). In the following consideration we assume the following
structure of the self-energy correction: 
\begin{equation}
\mathbf{\Sigma }_{k,\Lambda }=\Sigma _{k,\Lambda }^{(1)}\left( 
\begin{array}{cc}
1 & 1 \\ 
1 & 1%
\end{array}%
\right) +\Sigma _{k,\Lambda }^{(2)}\left( 
\begin{array}{cc}
1 & 1 \\ 
1 & 0%
\end{array}%
\right) ,  \label{SE_struct}
\end{equation}%
which was obtained in the ladder approximation in Ref. \cite{Our1PI} and
justified self-consistently below. With this assumption, the Green function $%
\mathbb{G}_{k,\Lambda }$ can be represented in the form%
\begin{equation}
\mathbb{G}_{k,\Lambda }=G_{k,\Lambda }\left( 
\begin{array}{cc}
1 & 0 \\ 
0 & 0%
\end{array}%
\right) +\frac{G_{\mathrm{loc},\nu }}{1-G_{\mathrm{loc},\nu }\Sigma
_{k,\Lambda }^{(2)}}\left( 
\begin{array}{cc}
-1 & 1 \\ 
1 & -1%
\end{array}%
\right)  \label{G_decompose}
\end{equation}%
where $G_{k,\Lambda }$ is given by the Eq. (\ref{G_iffl1}) with $\Sigma
_{k,\Lambda }=\Sigma _{k,\Lambda }^{(1)}+\Sigma _{k,\Lambda }^{(2)}$. For
the single-scale propagator we obtain%
\begin{equation}
\mathbb{S}_{k,\Lambda }=\left( 
\begin{array}{cc}
S_{k,\Lambda } & 0 \\ 
0 & 0%
\end{array}%
\right) ,  \label{S_struct}
\end{equation}%
where $S_{k,\Lambda }$ is identical to the single-scale propagator (\ref%
{SkL_iffl}), considered in Ref. \cite{ifflow}.

Considering the $\mathbf{\Sigma }_{k,\Lambda }^{11}=\Sigma _{k,\Lambda
}\equiv \Sigma _{k,\Lambda }^{(1)}+\Sigma _{k,\Lambda }^{(2)}$ component of
the self-energy and $\mathbb{V}_{\Lambda }^{1111}(k_{1},k_{2};k_{3},k_{4})$
component of the vertex, the first term in Eq. (\ref{G_decompose}) yields
the equations of the DMF$^{2}$RG approach, while the second term yields zero
in the assumption that the vertex keeps its structure (\ref{V_struct}). We
have to verify however, that the Eqs. (\ref{V_struct}) and (\ref{S_struct})
are preserved by the considering approach. Let consider first the
self-energy. Assuming fulfillment of the ansatz for the vertices (\ref%
{V_struct}) and using the result (\ref{S_struct}) for the single-scale
propagator, we find $\mathbf{\Sigma }_{k,\Lambda }^{11}=\mathbf{\Sigma }%
_{k,\Lambda }^{12}=\mathbf{\Sigma }_{k,\Lambda }^{21}$. This implies
fulfillment of (\ref{S_struct}). Let now verify fulfillment of the vertex
ansatz (\ref{V_struct}). Starting with this ansatz and (\ref{S_struct}), we
find that the first term in Eq. (\ref{G_decompose}) yields fulfillment of
the first two lines of (\ref{V_struct}). For the vertices $\mathbb{V}%
_{\Lambda }^{1211},$ $\mathbb{V}_{\Lambda }^{2111},$ $\mathbb{V}_{\Lambda
}^{1121},\mathbb{V}_{\Lambda }^{1112}$ the second term involves vertices
with two indices $"2",$ like $\mathbb{V}_{\Lambda }^{1212}$ etc, which are
however canceled because of the structure (\ref{G_decompose}) of the Green
function and single-scale propagator (\ref{S_struct}). Therefore, we find
that the representations, given by the first two lines of Eq. (\ref{V_struct}%
) and Eqs. (\ref{G_decompose}) and (\ref{S_struct}) keep their form during
the flow, and the considering approach appears to be equivalent to the DMF$%
^{2}$RG approach.

\section{The dual fermion approach}

The dual fermion approach of Refs. \cite{DF1,DF2,DF3,DF4} can be
conveniently formulated by splitting an effective interaction of the lattice
theory (see, e.g. Ref. \cite{Salmhofer}) 
\begin{eqnarray}
\mathcal{V}[\eta ,\eta ^{+}] &:&=-\ln \int d[c,c^{+}]\exp \left\{
\sum\limits_{k,\sigma }G_{0k}^{-1}\left( c_{k\sigma }^{+}+\eta _{k\sigma
}^{+})(c_{k\sigma }+\eta _{k\sigma }\right) \right.  \notag \\
&&\left. -\sum\limits_{i,\sigma }\int d\tau H_{\mathrm{int}}[c_{i\sigma
},c_{i\sigma }^{+}]\right\}  \notag \\
&=&-\ln Z[G_{0k}^{-1}\eta _{k\sigma },G_{0k}^{-1}\eta _{k\sigma }^{+}]-\eta
_{k\sigma }^{+}G_{0k}^{-1}\eta _{k\sigma },  \label{Vs}
\end{eqnarray}%
without performing its Legendre transformation. The expansion of $\mathcal{V}%
[\eta ,\eta ^{+}]$ in source fields generates connected (in general,
one-particle reducible) Green functions, amputated by the non-interacting
Green functions of the lattice theory $G_{0k}$. Application of covariation
splitting formula to the Eq. (\ref{Vs}) yields the relation between the
lattice and dual effective interactions\cite{MySixpt} 
\begin{equation}
\mathcal{V}[\eta ,\eta ^{+}]=\widehat{\mathcal{V}}[\widehat{\eta },\widehat{%
\eta }^{+}]+\sum_{k,\sigma }\eta _{k,\sigma }^{+}\frac{\Sigma _{\mathrm{loc}%
}(i\nu _{n})}{1-\Sigma _{\mathrm{loc}}(i\nu _{n})G_{0k}}\eta _{k,\sigma }
\label{VVRel}
\end{equation}%
where the effective interaction for the dual theory is defined by%
\begin{eqnarray}
\widehat{\mathcal{V}}[\widehat{\eta },\widehat{\eta }^{+}] &=&-\ln \int D[%
\widetilde{c},\widetilde{c}^{+}]e^{\sum\limits_{k,\sigma }\widetilde{%
\mathcal{G}}_{k}^{-1}\left( \widetilde{c}_{k,\sigma }^{+}-\widehat{\eta }%
_{k,\sigma }^{+}\right) \left( \widetilde{c}_{k,\sigma }-\widehat{\eta }%
_{k,\sigma }\right) -\widehat{\mathcal{V}}_{\mathrm{DMFT}}[\widetilde{c}^{+},%
\widetilde{c}]},  \label{Vez1} \\
\widehat{\eta }_{k\sigma } &=&\eta _{k\sigma }/\left[ 1-\Sigma _{\mathrm{loc}%
}(i\nu _{n})G_{0k}\right] ,  \notag
\end{eqnarray}%
as in previous Section we have introduced $\widetilde{\mathcal{G}}_{k}=%
\mathcal{G}_{k}-G_{\mathrm{loc}}$.

To introduce $\Lambda $-dependence of effective interaction, we replace
again, similarly to the previous Section, $\mathcal{G}_{k}\rightarrow 
\mathcal{G}_{k,\Lambda }$, see e.g. Eqs. (\ref{G_L_ifflow}) or Eqs. (\ref%
{G_L_sum}) and (\ref{G_tilde}). To generalize the relation (\ref{VVRel}) to
arbitrary $\Lambda ,$ we also replace $G_{0k}\rightarrow G_{0k,\Lambda }:=[%
\mathcal{G}_{k,\Lambda }^{-1}+\Sigma _{\mathrm{loc}}(i\nu _{n})]^{-1}.$ The
Eq. (\ref{VVRel}) then reads 
\begin{eqnarray}
\mathcal{V}_{\Lambda }[\eta ,\eta ^{+}] &=&\widehat{\mathcal{V}}_{\Lambda }[%
\widehat{\eta }_{\Lambda },\widehat{\eta }_{\Lambda }^{+}]+\sum_{k,\sigma }%
\widehat{\eta }_{k,\Lambda ,\sigma }^{+}\frac{\Sigma _{\mathrm{loc}}(i\nu
_{n})}{1+\mathcal{G}_{k,\Lambda }\Sigma _{\mathrm{loc}}(i\nu _{n})}\widehat{%
\eta }_{k,\Lambda ,\sigma }^{+},  \label{VVren} \\
\widehat{\eta }_{k,\Lambda ,\sigma } &=&\eta _{k\sigma }[1+\Sigma _{\mathrm{%
loc}}(i\nu _{n})\mathcal{G}_{k,\Lambda }]  \notag
\end{eqnarray}%
This allows us to perform \textit{consistent} renormalization of the lattice
and dual theory, keeping, in particular unchanged form of the relation
between the dual $\Sigma _{\mathrm{d}}(k,\Lambda )$ and lattice $\Sigma
_{k,\Lambda }$ self-energies\cite{DF1,MySixpt}: 
\begin{equation}
\Sigma _{k,\Lambda }=\frac{\Sigma _{\mathrm{d}}(k,\Lambda )}{1+G_{\mathrm{loc%
}}(i\nu _{n})\Sigma _{\mathrm{d}}(k,\Lambda )}+\Sigma _{\mathrm{loc}}(i\nu
_{n}).  \label{relSigma}
\end{equation}%
Note that the alternative way of keeping this relation is to relate the flow
in the dual space directly to the flow in the real space, as it is done in
Ref. \cite{OurRG}.

The renormalization of the dual fermion effective interaction $\widehat{%
\mathcal{V}}_{\Lambda }[\widehat{\eta },\widehat{\eta }^{+}]$ can be
performed in the standard way. The Polchinskii equation for $\widehat{%
\mathcal{V}}_{\Lambda }$ reads: 
\begin{equation*}
\partial _{\Lambda }\widehat{\mathcal{V}}_{\Lambda }[\widehat{\eta },%
\widehat{\eta }^{+}]=-\Delta _{\partial _{\Lambda }\widetilde{G}_{k,\Lambda
}}\widehat{\mathcal{V}}_{\Lambda }+\Delta _{\partial _{\Lambda }\widetilde{G}%
_{k,\Lambda }}^{12}\widehat{\mathcal{V}}_{\Lambda }^{(1)}\widehat{\mathcal{V}%
}_{\Lambda }^{(2)}
\end{equation*}%
Note that in the latter equation the $\Lambda $-derivative does not act on
the source fields $\widehat{\eta },\widehat{\eta }^{+}$, which $\Lambda $%
-dependent values are substituted into the resulting effective interaction.
The latter $\Lambda $-dependence determines the flow of the lattice
effective interaction (\ref{VVren}) according to%
\begin{eqnarray}
\partial _{\Lambda }\mathcal{V}_{\Lambda }[\eta ,\eta ^{+}] &=&\partial
_{\Lambda }\widehat{\mathcal{V}}_{\Lambda }[\widehat{\eta }_{\Lambda },%
\widehat{\eta }_{\Lambda }^{+}]+\left\{ \eta \frac{\delta \widehat{\mathcal{V%
}}_{\Lambda }}{\delta \widehat{\eta }_{\Lambda }}+\frac{\delta \widehat{%
\mathcal{V}}_{\Lambda }}{\delta \widehat{\eta }_{\Lambda }^{+}}\eta
^{+}\right\} (\Sigma _{\mathrm{loc}}\partial _{\Lambda }G_{k,\Lambda }) 
\notag \\
&&+\sum_{k,\sigma }\eta _{k,\sigma }^{+}\Sigma _{\mathrm{loc}}^{2}(i\nu
_{n})(\partial _{\Lambda }\mathcal{G}_{k,\Lambda })\eta _{k,\sigma }
\end{eqnarray}%
Assuming 
\begin{eqnarray}
\mathcal{V}_{\Lambda }[\eta ,\eta ^{+}] &=&\sum \overline{V}_{n,\Lambda
}\eta _{1}^{+}..\eta _{n/2}^{+}\eta _{n/2+1}..\eta _{n}  \notag \\
\widehat{\mathcal{V}}_{\Lambda }[\widehat{\eta },\widehat{\eta }^{+}]
&=&\sum \overline{v}_{n,\Lambda }\widehat{\eta }_{1}^{+}..\widehat{\eta }%
_{n/2}^{+}\widehat{\eta }_{n/2+1}..\widehat{\eta }_{n}
\end{eqnarray}%
this yields the standard relation between the lattice and dual two-point,
Eq. (\ref{relSigma}) and higher-order vertices%
\begin{equation}
\overline{V}_{n,\Lambda }=\overline{v}_{n,\Lambda }\prod\limits_{i=1}^{n}%
\left[ 1+\Sigma _{\mathrm{loc}}(i\nu _{i})\mathcal{G}_{k_{i},\Lambda }\right]
\ \ (n>2).  \label{VRel}
\end{equation}%
The latter relation accounts for the effect of the missing local self-energy
insertions in the effective interaction $\widehat{\mathcal{V}}_{\mathrm{DMFT}%
}[\widetilde{c}^{+},\widetilde{c}],$ which determines $\widehat{\mathcal{V}}%
_{\Lambda }[\widehat{\eta },\widehat{\eta }^{+}]$ according to the Eq. (\ref%
{Vez1}).

The Legendre transformation of $\widehat{\mathcal{V}}_{\Lambda }$ can be
also performed in the standard way. The resulting 1PI (with respect to $%
\widetilde{G}_{k}$) fRG equations for the fully amputated vertex $v_{\Lambda
}=\overline{v}_{\Lambda }\prod\nolimits_{i=1}^{4}\widetilde{\mathcal{G}}%
_{k_{i},\Lambda }\widetilde{G}_{k_{i},\Lambda }^{-1}$ $(\overline{v}%
_{\Lambda }\equiv \overline{v}_{4,\Lambda })$ read 
\begin{eqnarray}
\frac{d\Sigma _{\mathrm{d}}}{d\Lambda } &=&v_{\Lambda }\circ S_{\Lambda }
\label{OneLoopDF1} \\
\frac{dv_{\Lambda }}{d\Lambda } &=&v_{\Lambda }\circ (\widetilde{G}_{\Lambda
}\circ S_{\Lambda }+S_{\Lambda }\circ \widetilde{G}_{\Lambda })\circ
v_{\Lambda }  \label{OneLoopDF}
\end{eqnarray}%
where 
\begin{eqnarray}
\widetilde{G}_{k,\Lambda } &=&\widetilde{\mathcal{G}}_{k,\Lambda }/[1-\Sigma
_{\mathrm{d}}(k,\Lambda )\widetilde{\mathcal{G}}_{k,\Lambda }],  \label{G_DF}
\\
v_{\Lambda } &=&\overline{v}_{\Lambda }\prod\limits_{i=1}^{4}(1-\Sigma _{%
\mathrm{d}}(k_{i},\Lambda )\widetilde{\mathcal{G}}_{k_{i},\Lambda }),
\label{vv}
\end{eqnarray}%
and 
\begin{equation}
S_{k,\Lambda }=\left. \frac{d\widetilde{G}_{k,\Lambda }}{d\Lambda }%
\right\vert _{\Sigma _{\mathrm{d}}=const}=-(\mathcal{G}_{k}^{-1}-G_{\mathrm{%
loc}}^{-1})\mathcal{G}_{k,\Lambda }^{2}\frac{\partial f(k,\Lambda )}{%
\partial \Lambda }\frac{1}{[1-\Sigma _{\mathrm{d}}(k,\Lambda )\widetilde{%
\mathcal{G}}_{k,\Lambda }]^{2}}  \label{S_DF2}
\end{equation}%
for the choice (\ref{G_L_ifflow}) and 
\begin{equation}
S_{k,\Lambda }=\left. \frac{d\widetilde{G}_{k,\Lambda }}{d\Lambda }%
\right\vert _{\Sigma _{\mathrm{d}}=const}=-(\mathcal{G}_{k}-G_{\mathrm{loc}})%
\frac{\partial f(k,\Lambda )}{\partial \Lambda }\frac{1}{[1-\Sigma _{\mathrm{%
d}}(k,\Lambda )\widetilde{\mathcal{G}}_{k,\Lambda }]^{2}}  \notag
\end{equation}%
for the choice (\ref{G_L_sum}). The initial condition reads $\Sigma _{%
\mathrm{d}}=0,$ $V_{\Lambda }=\Gamma _{\mathrm{loc}}$. As discussed above,
these equation neglect the local six-point vertex, which is one-particle
reducible with respect to the local Green function.

To compare the equations (\ref{OneLoopDF1}) and (\ref{OneLoopDF}) to those
of the DMF$^{2}$RG approach, we consider again cutoff dependence (\ref%
{G_L_ifflow}). The vertices $v_{n,\Lambda }$ with $n>2$ can be related to
the corresponding vertices $V_{n,\Lambda }$ of the DMF$^{2}$RG approach by
amputating $\overline{V}_{n,\Lambda }$ by the respective full Green
functions and using the relations (\ref{VRel}) and (\ref{vv}): $V_{n,\Lambda
}=\overline{V}_{n,\Lambda }\prod\nolimits_{i=1}^{n}G_{0k_{i},\Lambda
}G_{k_{i},\Lambda }^{-1}=v_{n,\Lambda }\prod\nolimits_{i=1}^{n}\{[1+\Sigma _{%
\mathrm{loc}}(i\nu _{i})\mathcal{G}_{k_{i},\Lambda }][1-\Sigma
(k_{i},\Lambda )G_{0k_{i},\Lambda })]/[1-\Sigma _{\mathrm{d}}(k_{i},\Lambda )%
\widetilde{\mathcal{G}}_{k_{i},\Lambda }]\}$. It can be verified that this
factor cancels exactly the difference between the single-scale propagator
and Green functions in the dual fermion approach [Eqs. (\ref{G_DF}) and (\ref%
{S_DF2})] and the DMF$^{2}$RG approach [Eqs. (\ref{G_iffl1}) and (\ref%
{SkL_iffl})], see Ref. \cite{OurRG}. The corresponding equations differ
however because of the $\Lambda $-derivative of the corresponding factors%
\cite{OurRG}.

\section{Cutoff schemes and self-consistency}

Here we compare different cutoff schemes and analyze their applicability to
the renormalization-group treatment, discussed in previous Sections.

We start with the simple momentum cutoff%
\begin{equation}
f(k,\Lambda )=\theta (|\varepsilon _{\mathbf{k}}|-\Lambda )  \label{GL}
\end{equation}%
Being combined with the Eq. (\ref{G_L_sum}), the choice (\ref{GL}) has
simple physical meaning: we put the Green function equal to the local Green
function inside the shell $|\varepsilon _{\mathbf{k}}|<\Lambda $ and equal
to the non-local function outside this shell. This cutoff, however, does not
preserve the important property of vanishing of average of $\widetilde{G}%
_{\Lambda }$ over momentum space during the flow: 
\begin{equation}
\sum_{\mathbf{k}}\widetilde{G}_{\Lambda }(\mathbf{k},i\nu _{n})=\sum_{%
\mathbf{k:}|\varepsilon _{\mathbf{k}}|>\Lambda }[G(k,i\nu _{n})-G_{\mathrm{%
loc}}(i\nu _{n})]\neq 0 \ \ (\Lambda >0)
\end{equation}

The other possible choices are the `interaction flow' cutoff%
\begin{equation}
f(k,\Lambda )=1-\Lambda
\end{equation}%
and frequency cutoff by C. Husemann and M. Salmhofer, Ref. \cite{Salmhofer1},%
\begin{equation}
f(k,\Lambda )=\frac{\nu _{n}^{2}}{\nu _{n}^{2}+\Lambda ^{2}/(1-\Lambda )^{2}}
\end{equation}%
which allow to flow from the theory, non-locally non-interacting ($\Lambda
=1 $) to fully interacting one ($\Lambda =0$). These two cutoffs preserve
the local part of the Green function, provided that the decomposition (\ref%
{G_L_sum}) is used. The possible disadvantage of the latter two cutoffs is
the large computational effort: since one can not project momenta to the
Fermi surface, one has to deal with many `patches' in the whole Brilloin
zone.

Finally, we comment on the effect of the self-consistency. In the dual
fermion approach two ingridients of self-consistent procedure were used. The
first one is obtaining the self-consistent self-energy using the diagram
series in auxiliary space. This self-consistency is fully implemented in the
discussed approaches via flowing self-energy. At the same time, second step
(the so-called external self-consistency) requires adjusting the initial
local problem according to the local part of the obtained self-energy.
Similar procedure can be applied to the approaches, considered in the
present paper. This type of the self-consistency is expected to be important
at relatively strong coupling, where it increases the resulting self-energy,
making it more "insulating\textquotedblright , see, e.g. Ref. \cite{Our1PI}.

\section{Conclusion}

We have considered the application of functional renormalization-group
approach to strongly-correlated electronic systems within the one-particle
irreducible approach with respect to the local Green functions (1PI-LGF) and
the dual fermion approach. Both mentioned approaches allow for consistent
renormalization; the dual fermion approach is expected to be applicable if
the one-particle reducible (with respect to the local Green functions)
vertices of sixth- and higher orders are small, while the 1PI-LGF approach
assumes smalleness of one-particle irreducible vertices.

Further numerical investigations of the validity of these assumptions, as
well as comparison of the results of the presented approaches to the flow
from infinite to finite dimensions \cite{ifflow} to be performed.

\textit{Acknowledgements. }The author is grateful to G. Rohringer, C.
Taranto, A. Toschi, K. Held, C. Honerkamp, N. Wentzell, S. Andergassen, and
A. I. Lichtenstein for stimulating discussions. The work is supported by the
grant of Dynasty foundation.

\end{document}